\documentclass[%
reprint,
 showkeys,
 amsmath,amssymb,
 aps,
 pra,
]{revtex4-1}
\usepackage{graphics,epsfig}
\usepackage[]{graphicx}
\usepackage{latexsym}
\usepackage{amsmath, amsthm, amsfonts, amssymb}
\usepackage{verbatim}
\usepackage{graphicx}% Include figure files
\usepackage{dcolumn}% Align table columns on decimal point
\usepackage{bm}

\begin{document}

\preprint{APS/123-QED}

\newcommand{\nd}{\noindent}
\newtheorem{theo}{Theorem}[section]
\newtheorem{definition}[theo]{Definition}
\newtheorem{lem}[theo]{Lemma}
\newtheorem{prop}[theo]{Proposition}
\newtheorem{coro}[theo]{Corollary}
\newtheorem{exam}[theo]{Example}
\newtheorem{rema}[theo]{Remark}
\newtheorem{example}[theo]{Example}
\newtheorem{principle}[theo]{Principle}
\newcommand{\ninv}{\mathord{\sim}} %involutive negation
\newtheorem{axiom}[theo]{Axiom}
%\numberwithin{equation}{subsection}

\newcommand{\be}{\begin{equation}}
\newcommand{\ee}{\end{equation}}
\newcommand{\ben}{\begin{eqnarray}}
\newcommand{\een}{\end{eqnarray}}
\newcommand{\nn}{\nonumber \\}
\newcommand{\ii}{\'{\i}}
\newcommand{\pp}{\prime}

\newcommand{\tc}{\textcolor{red}}
\newcommand{\tb}{\textcolor{blue}}
\newcommand{\tg}{\textcolor{green}}

\title{Generalizing entanglement via informational invariance for arbitrary statistical theories}% Force line breaks with \\

\author{F. Holik$^{1,\,2}$}
\affiliation{$^{1}$Departamento de Matem\'{a}tica - Ciclo B\'{a}sico Com\'{u}n\\
Universidad de Buenos Aires - Pabell\'{o}n III, Ciudad
Universitaria \\ Buenos Aires, Argentina}

\affiliation{$^{2}$ Postdoctoral Fellow of CONICET-IFLP}

\author{C. Massri$^{3}$}
\affiliation{
 $^{3}$Departamento de Matem\'{a}tica de la Universidad de Buenos Aires}

\author{A. Plastino$^{4,\,5}$}
\affiliation{$^{4}$ National University La Plata \& CONICET
IFLP-CCT, C.C. 727 - 1900 La Plata, Argentina}

\affiliation{ $^{5}$ Universitat de les Illes Balears and
IFISC-CSIC, 07122 Palma de Mallorca, Spain}

\date{\today}% It is always \today, today,

\begin{abstract}
\noindent Given an arbitrary statistical theory, different from
quantum mechanics, how to decide which are the nonclassical
correlations? We present a formal framework which allows for a
definition of nonclassical correlations in such theories,
alternative to the current one. This enables one to formulate
extrapolations of some important quantum mechanical features via
adequate  extensions of ``reciprocal" maps relating states of a
system with states of its subsystems. These extended maps permit one
to generalize i) separability measures to any arbitrary statistical
model as well as ii) previous entanglement criteria. The standard
definition of entanglement becomes just a particular case of the
ensuing, more general notion.
\begin{description}
\item[PACS numbers]
\textbf{03.65.Ud}
\end{description}
\end{abstract}

\pacs{Valid PACS appear here}% PACS, the Physics and Astronomy
%Use showkeys class option if keyword
\maketitle

\bibliography{pom}

\begin{thebibliography}{10}

\bibitem{barnum} H. Barnum, J. Barrett, L. O. Clark,
M. Leifer, R. Spekkens, N. Stepanik, A. Wilce and R. Wilke, 2010
New J. Phys. {\bf 12}, 033024 (2010), and references therein.

\bibitem{BEN} I. Bengtsson, K. Zyczkowski, Geometry of Quantum States:
An Introduction to Quantum Entanglement, Cambridge Univ. Press,
Cambridge, 2006.

\bibitem{Schro1} E. Schr\"{o}dinger, Proc. Cambridge Philos. Soc. 31, 555 (1935).

\bibitem{Schro2} E. Schr\"{o}dinger, Proc. Cambridge Philos. Soc. 32, 446 (1936).

\bibitem{EPR} A. Einstein, B. Podolski, and N. Rosen, Phys. Rev. 47, 777 (1935)

\bibitem{ReviewHorodeki2009} R. Horodeki, P. Horodki, M. Horodeki, and K. Horodeki, Rev. Mod.
Phys. \textbf{81}, 865 (2009).

\bibitem{Holik-Plastino-2011a} F. Holik and A. Plastino, Phys. Rev. A
\textbf{84}, 062327 (2011).

\bibitem{Barnum-Wilce-2009} H. Barnum and A. Wilce, Electronic Notes in Theoretical Computer Science
Volume \textbf{270}, Issue 1, Pages 3-15,(2011).

\bibitem{Barret-2007} J. Barrett, Phys. Rev. A 75, 032304 (2007).

\bibitem{Perinotti-2010} G. Chiribella, G. M. D'Ariano, and P. Perinotti, Phys. Rev. A
81, 062348 (2010).

\bibitem{Barnum-Entropy} H. Barnum, J. Barrett, L. Orloff Clark, M. Leifer, R.
Spekkens, N. Stepanik, A. Wilce, and R. Wilke, New J. Phys. 12,
033024 (2010).

\bibitem{Entropy-generalized-II} A. J. Short and S. Wehner, New J. Phys. 12, 033023 (2010).

\bibitem{Hanggi-2009} E. H\"{a}nggi, R. Renner and S. Wolf, arXiv:0906.4760 (2009).

\bibitem{Pawlowski-2009} M. Pawlowski, T. Paterek, D. Kaszlikowski, V. Scarani, A. Winter and
M. Zukowski, arXiv:0905.2292 (2009).

\bibitem{Barnum-Dahlsten-Leifer-Toner-2008} H. Barnum, O. Dahlsten, M. Leifer and B. Toner, IEEE ITW pp. 386�90
(2008).

\bibitem{Buhram-2006} H. Buhrman, M. Christandl, F. Unger, S. Wehner and A. Winter, Proc. R. Soc. A 462
1919�32 (2006).

\bibitem{Gisin-2006} A. Short, N. Gisin and S. Popescu, Quantum Inf. Process. \textbf{5} 1573 (2006).

\bibitem{VerSteeg-Wehner-2009} G. Ver Steeg and S. Wehner, Quantum Inf. Comput. \textbf{9}
801 (2009)

\bibitem{vanDam} W. van Dam, arXiv:quant-ph/0501159 (2005).

\bibitem{Wolf} S. Wolf and J. Wullschleger, arXiv:quant-ph/0508233 (2005).

\bibitem{Perinotti-2011} G. Chiribella, G. M. D'Ariano, and P. Perinotti, Phys. Rev. A
84, 012311 (2011).

\bibitem{Barnum-Wilce-2010} H. Barnum, R. Duncan and A Wilce, arXiv:1004.2920v1 (2010)

\bibitem{Barnum-PRL} H. Barnum, J. Barrett, M. Leifer, and A.Wilce, Phys. Rev. Lett.
99, 240501 (2007).

\bibitem{MielnikGQS} B. Mielnik, Commun. math. Phys. \textbf{9} (1968) 55-80

\bibitem{MielnikTF} B. Mielnik, Commun. math. Phys. \textbf{15} (1969) 1-46

\bibitem{MielnikGQM} B. Mielnik, Commun. math. Phys. \textbf{37} (1974) 221-256

\bibitem{Ferrie-2011} C. Ferrie, Rep. Prog. Phys. \textbf{74} 116001
(2011).

\bibitem{Holik-Massri-Plastino-Zuberman} F. Holik, C. Massri, A.
Plastino and L. Zuberman, Int. Jour. Theo. Phys., in press (2012).

\bibitem{wbook}  C. Zachos, D. B. Fairlie, T. L. Curtright, Eds., {\it Quantum mechanics in phase
space}(World scientific, Singapore, 2005).

\bibitem{Effects-Biology} M. A. Martin-Delgado, Scientific Reports \textbf{2} 302 DOI: 10.1038/srep00302

\bibitem{DVB10} B. Dakic, V. Vedral, C. Brukner, Phys. Rev. Lett. {\bf 105} (2010) 190502.

\bibitem{OZ02} H. Ollivier, W.H. Zurek, Phys. Rev. Lett {\bf 88} (2001) 017901.

\bibitem{CABMPW10} D. Cavalcanti, L. Aolita, S. Boixo, K. Modi, M. Piani, A. Winter Phys. Rev. A {\bf83} (2011) 032324.

\bibitem{batle}  J. Batle, A. R. Plastino, A. Plastino, M. Casas,
J.Phys. A {\bf 44} (2011) 503304.

\bibitem{ana}  A.P. Majtey,  A.R. Plastino,  A. Plastino, {\it  New features
of quantum discord uncovered by q-entropies},
 Physica A (2011)in Press.

\bibitem{HV01} L. Henderson and V. Vedral, J. Phys. A {\bf 34} (2001) 6899.

\bibitem{LLZ07} N. Li, S. Luo, Z. Zhang, J. Phys. A {\bf 40} (2007) 11361.

\bibitem{Perinotti-2011} P. Perinotti, \textit{Phys. Rev. Lett.}, vol. \textbf{108}, Issue \textbf{12}, (2012)
120502.

\bibitem{Barnum-Toner} H. Barnum, O. C. O. Dahlsten, M. Leifer, and B. Toner, in
Information Theory Workshop, 2008, pp. 386 -390, (2008).

\bibitem{Beltrametti.Varadarajan-2000} E. Beltrametti, S. Bugajski
and V. Varadarajan, J. Math. Phys. \textbf{41} (2000)

\bibitem{Abascal2007} I. S. Abascal and G. Bj\"{o}rk, Phys. Rev. A \textbf{75}, 062317 (2007).

\bibitem{Bjork-2007a} C. Kothe and G. Bj\"{o}rk, Phys.Rev.A \textbf{75}, 012336 (2007).

\bibitem{Bjork-2007b} C. Kothe, I. Sainz, and G. Bj\"{o}rk,J. Phys.: Conf. Ser. \textbf{84}, 012010
(2007).

\bibitem{Zhangh-2008} C. J. Zhang, Y. S. Zhang, S. Zhang, and G. C. Guo, Phys. Rev.
A \textbf{77}, 060301 (2008).

\bibitem{Ent95} J. Schlienz and G. Mahler, Phys. Rev. A \textbf{52}, (1995) 4396

\bibitem{foulis} D. Foulis, M. K. Bennett,  Found. Phys. {\bf 24}
(1994) 1331�1352.

\bibitem{bush} P. Busch, P. Lahti, P. Mittlestaedt, {\it The Quantum Theory of Measurement} (Springer-Verlag,
Berlin, 1991).

\bibitem{Holik-Plastino-2011b} F. Holik and A. Plastino, \textit{J. Math. Phys.} \textbf{53}, 073301 (2012).

\bibitem{stulpe2001} W. Stulpe and  M. Swat, Found. of Phys. Lett.
{\bf 14}, 285 (2001).

\bibitem{BarnumArXiv} H. Barnum, R. Duncan, and A. Wilce, e-print arXiv:1004.2920v1.

\bibitem{Gudder-StatisticalMethods} S. P. Gudder, \textit{Stochastic
Methods in Quantum Mechanics} North Holland, New York - Oxford
(1979)

\bibitem{gudderlibro78} S. P. Gudder, in {\it Mathematical Foundations
of Quantum Theory}, A. R. Marlow, ed. (Academic, New York, 1978)

\bibitem{wilce} A. Wilce, {\it Quantum Logic and Probability Theory},
The Stanford Encyclopedia of Philosophy (Spring 2009 Edition),
Edward N. Zalta (ed.), URL =
http://plato.stanford.edu/archives/spr2009/entries/qt-quantlog/.
Archive edition: Spring 2009.

\bibitem{Ferraro-discord-2010} A. Ferraro, L. Aolita, D. Cavalcanti,
F. M. Cuchietti and A. Ac\'{i}n, Phys. Rev. A \textbf{81}, 052318
(2010).

\bibitem {Werner} R. Werner, Phys. Rev. A \textbf{40}, (1989) 4277-4281

\end{thebibliography}

\section{Introduction}

\nd Quantum mechanics can be regarded as an extension of the
classical probability calculus that allows for random variables that
are not simultaneously measurable \cite{barnum}. Working from this
peculiar perspective, it can be shown that many phenomena usually
considered as typically quantal, like quantum no-cloning and
no-broadcasting theorems, the trade-off between state disturbance
and measurement, and the existence and basic properties of entangled
states, are in fact generic features of non-classical probabilistic
theories that verify a basic {\it non-signaling} constraint
\cite{barnum}. This is the point of departure of our present
considerations.

\nd In particular, entanglement \cite{BEN} is conventionally viewed
as the most emblematic expression of non-classicality.
Schr\"{o}dinger is widely quoted stating that ``entanglement is
\emph{the} characteristic trait of quantum mechanics"
\cite{Schro1,Schro2,EPR}. Indeed, characterizing entanglement has
become one of the most important current tasks of physics
\cite{ReviewHorodeki2009}, with a host of possible technological
applications. An entanglement criterion based on {\it geometrical}
properties of entanglement has been recently presented in
\cite{Holik-Plastino-2011a}. These geometrical features of
entanglement will be employed here to extrapolate many
entanglement's properties to {\it arbitrary probabilistic theories}.
This is done by recourse to an essential mathematical ingredient,
the so-called Convex Operational Model (COM) approach. The COM
approach is founded on geometrical properties of a special convex
set, that containing all the states of an arbitrary statistical
theory \cite{Barnum-Wilce-2009,Barnum-Wilce-2010} (see also
\cite{Perinotti-2010,Perinotti-2011,Barnum-Entropy,Barnum-PRL,Barnum-Toner,BarnumArXiv,Barret-2007,Entropy-generalized-II}).
The COM approach has its roots in operational theories and has been
shown to be useful to generalize many quantum mechanical notions
mentioned above, such as teleportation protocols, no broadcasting,
and no cloning theorems
\cite{Barnum-Wilce-2009,Barnum-Wilce-2010,Barnum-PRL}. The
geometrical approach based on convex sets can also be seen as a
framework in which non-linear theories which generalize quantum
mechanics, can be included, studied, and compared with it
\cite{MielnikGQM,MielnikGQS,MielnikTF}. It is also important to
remark that an axiomatization independent (and equivalent to) the
von Neumann formalism can be given using the geometrical-operational
approach \cite{MielnikGQM,MielnikGQS,MielnikTF}.

\nd The importance of entanglement as a resource for measuring
classicality of a state has been highlighted in
\cite{Ferrie-2011}. Other measures of non-classicality exist, of
course. One of the most important is the negativity of the Wigner
function \cite{wbook}. Another important measure of
non-classicality -often found in quantum optics- has to do with
the properties of coherent states, i.e., a state will be
considered classical if it can be written as a mixture of coherent
sates (which satisfy a minimal violation of Heisenberg's
uncertainty principle). More recently, quantum discord (QD)
\cite{DVB10,OZ02,LLZ07,CABMPW10,batle,ana} has became another
measure of non-classicality. QD refers to important manifestations
of the quantumness of correlations in composite systems that are
different from those of entanglement-origin and  may be relevant
in quantum information technologies
\cite{DVB10,OZ02,CABMPW10,HV01}.

\vskip 3mm

\nd In this work, we restrict ourselves to entanglement (see
\cite{Perinotti-2011} for the QD case) and provide a
characterization of it using maps. We show that this
characterization can be generalized to arbitrary statistical models.
The issue has been studied, for example, in
\cite{Perinotti-2011,Barnum-Toner}. Our entanglement-extension
(based in \cite{Holik-Plastino-2011a}) allows for an alternative
approach, which provides a quite general characterization of non
classical correlations in arbitrary statistical models,
\emph{leaving the standard treatment as a particular case}.

\nd More explicitly, our characterization of entanglement is based
on the maps that relate states of the system with states of its
subsystems. In particular, following the generalization presented in
\cite{Beltrametti.Varadarajan-2000}, we define generalized partial
traces by imposing conditions on ($1$) \emph{morphisms between
extensions of convexity models} and ($2$) a \emph{special map} which
allows one to create the set of separable states given the available
states of two parties. Differently from the standard approach
\cite{Perinotti-2011,Barnum-Toner}, in our proposal the
characterization of non-classical correlations is based on maps.

\noindent The interlink between these two items is investigated,
and, via appeal to constructions presented in
\cite{Holik-Plastino-2011a}, we concoct a geometrical
characterization of entanglement in arbitrary COM's. Specifically,
we generalize the notion of \emph{informational invariance},
advanced in \cite{Holik-Plastino-2011a}. It is shown that this
characterization of entanglement lies at the heart of the
separability problem in any statistical theory, providing i) an
alternative visualization of it and ii) enriching the
convex/operational approach to QM
\cite{Beltrametti.Varadarajan-2000,MielnikGQM,MielnikGQS,MielnikTF}
(as well as to other statistical theories).

\vskip 3mm

\nd The alternative perspective presented in this work will allow
us to obtain, for  a canonical family of separability measures
(based on the Schlienz-Mahler one \cite{Ent95}), its most general
form. As a result, we will be able to construct a general
quantitative (and in many cases computable) measure of
non-classicality for arbitrary statistical theories, including non
linear generalizations of QM. This general characterization of a
vast family of entanglement measures will permit one to  compare
the behavior of measures of non-classical correlations in
different theories, and thus, to single out specifical features of
QM. Why is this of importance? The answer is given in,  for
example,  \cite{Barnum-Entropy} and \cite{Entropy-generalized-II}.
Several possible applications were envisaged in
\cite{Hanggi-2009,Barnum-Dahlsten-Leifer-Toner-2008,Pawlowski-2009,Buhram-2006,Gisin-2006,VerSteeg-Wehner-2009,vanDam,Wolf}.

\noindent Since our constructions and their implications are
formulated in the geometrical setting of the COM approach, they
could become applicable to many physical theories of interest. An
example of such theories are ``Popescu-Rohrlich" boxes
\cite{Barret-2007}. Our construction could also be applied to
quantum mechanics with a limited set of allowed measurements,
general $C^\ast$-algebraic theories, theories derived by relaxing
uncertainty relations, etc.

\noindent In principle, the scope of the generalization given by
the COM approach is not constrained to physical theories. It also
includes mathematical models of any statistical theory, provided
these theories satisfy very general requirements. Thus, the
generalization of quantum mechanical notions --\emph{and specially
measures of non-classical correlations}-- to arbitrary statistical
theories via the COM approach is a useful alternative tool for
extrapolating such notions to different fields of research. For
example, the influence of quantum effects and entanglement in
evolution was studied in a toy biological model based on a
Chaitin's idea \cite{Effects-Biology}. The study of more realistic
models may require rather sophisticated mathematical frameworks
for which the COM approach and the kind of generalization
presented in this work (as well as in others, for example
\cite{Barnum-PRL}), can be useful.

\vskip 1mm \nd In Section \ref{dos} we briefly recapitulate the
notion of quantum effects and in section \ref{s:COMpreliminaries} we
review the COM approach. Next, in section \ref{s:paperPRA}, we write
in a convenient form the main details of the geometrical structure
that underlies entanglement, as advanced in
\cite{Holik-Plastino-2011a}. By following
\cite{Beltrametti.Varadarajan-2000} we build in section
\ref{s:EntanglementGeneralization} a geometrical generalization of
the relevant structures, and discuss its application to the
development of generalized entanglement measures. Finally, in
section \ref{s:conclusions} some conclusions are drawn. An Appendix
on quantal effects is also provided.

\section{Quantal effects} \label{dos}

\nd An algebraic structure called an effect algebra has been
introduced for investigations in the foundations of quantum
mechanics \cite{foulis}. The elements of an effect algebra
$\mathcal{E}$ are called quantum effects and are very important
indeed for quantum statistics and quantum measurement theory
\cite{bush}. One may regard a quantum effect as an elementary yes-no
measurement that may be un-sharp or imprecise.

\nd Quantum effects are used to construct generalized quantum
measurements (or observables). The structure of an effect algebra is
given by a partially defined binary operation $\bigoplus$ that is
used to form a combination $a \bigoplus b$ of effects $a, b \in
\mathcal{E}$. The element $a \bigoplus b$ represents a statistical
combination of $a$ and $b$ whose probability of occurrence equals
the sum of the probabilities that $a$ and $b$ occur individually.
Usually, effect algebras  possess a convex structure. For example,
if $a$ is a quantum effect and $\lambda \in [0, 1]$, then $\lambda
a$ represents the effect $a$ attenuated by a factor of $\lambda$.
Then, $\lambda a \bigoplus (1 - \lambda) b$ is a generalized convex
combination that can be constructed in practice. If a quantum system
$\cal{S}$ is represented by a Hilbert space $\cal{H}$, then a
self-adjoint operator $\hat A$ such that $0 \le \hat A \le 1$
corresponds to an effect for $\cal{S}$ \cite{foulis}. For more
details, see Appendix A.

\section{COM's preliminaries}\label{s:COMpreliminaries}

\nd Following \cite{Barnum-PRL}, we now review elementary
COM-notions. The aim of this formalism is to model general
statistical or operational theories. Any statistical theory has a
set of states $\omega \in \Omega$ and a set of observables.

\nd  It is reasonable to postulate that the set $\Omega$ is
convex, because the mixture of two states in any statistical
theory ought to yield a new state. For the convex set $\Omega$ one
should then associate probabilities to any observable $a$. This
entails that one must define a probability $a(\omega)\in[0,1]$ for
any state $\omega\in\Omega$. Usually, any observable is an affine
functional belonging to a space $A(\Omega)$ (the space of all
affine functionals). It is also assumed that there exists a
unitary observable $u$ such that $u(\omega)=1$ for all
$\omega\in\Omega$ and (in analogy with the quantum case, in which
they form an ordered space), the set of all quantum effects (the
reader not familiarized with the concept is advised to look at
Appendix A) will be encountered in the interval $[0,u]$. A
measurement will be represented by a set of effects $\{a_{i}\}$
such that $\sum_{i}a_{i}= u$.

\nd $\Omega$ is then naturally embedded
($\omega\mapsto\hat{\omega}$) in the dual space $A(\Omega)^{\ast}$
as follows: $\hat{\omega}(a):=a(\omega)$. Call $V(\Omega)$  the
linear span of $\Omega$ in $A(\Omega)^{\ast}$. $\Omega$ will be
considered finite dimensional if and only if $V(\Omega)$ is finite
dimensional, and we restrict ourselves to such situation (and to
compact spaces). This implies that $\Omega$ will be the convex hull
of its extreme points, called pure states (for details see, for
example, \cite{Barnum-Toner,Barnum-Wilce-2009,Barnum-Wilce-2010}).
In a finite dimension $d$ {\it a system will be classical if and
only if it is a simplex, i.e., the convex hull of $d+1$ linearly
independent pure states}. It is a well known fact that in a simplex
a point may be expressed as a unique convex combination of its
extreme points, {\it a characteristic feature of classical theories
that no longer holds in a quantum one}.

\vskip 1mm \nd {\sf Summing up}, a COM may be regarded as a
triplet $(\mathbf{A},\mathbf{A}^{\ast},u_{\mathbf{A}})$, where
$\mathbf{A}$ is a finite dimensional vector space,
$\mathbf{A}^{\ast}$ its dual  and $u_{\mathbf{A}}\in\mathbf{A}$ is
a unit functional.

\vskip 1mm \nd For compound systems, if its components have state
spaces $\Omega_{A}$ and $\Omega_{B}$, let  $\Omega_{AB}$ denote the
joint state space. Under reasonable assumptions, it turns out
\cite{Barnum-PRL} that $\Omega_{AB}$ may be identified with a linear
span of $(V(\Omega_{A})\otimes V(\Omega_{B}))$. A maximal tensor
product state space $\Omega_{A}\otimes_{max}\Omega_{B}$ can be
defined as the one which contains all bilinear functionals
$\varphi:A(\Omega_{A})\times A(\Omega_{B})\longrightarrow\mathbb{R}$
such that $\varphi(a,b)\geq 0$ for all effects $a$ and $b$ and
$\varphi(u_{A},u_{B})=1$. The maximal tensor product state space has
the property of being the biggest set of states in
$(A(\Omega_{A})\otimes A(\Omega_{B}))^{\ast}$ which assigns
probabilities to all product- measurements.

\nd On the other hand, the minimal tensor product state space
$\Omega_{A}\otimes_{min}\Omega_{B}$ is defined as the one which is
formed by the convex hull of all product states. A product state is
a state of the form $\omega_{A}\otimes\omega_{B}$ such that
$\omega_{A}\otimes\omega_{B}(a,b)=\omega_{A}(a)\omega_{B}(b)$ for
all pairs $(a,b)\in A(\Omega_{A})\times A(\Omega_{B})$. The actual
set of states $\Omega_{AB}$ (to be called
$\Omega_{A}\otimes\Omega_{B}$ from now on) of a particular system
will satisfy
$\Omega_{A}\otimes_{min}\Omega_{B}\subseteq\Omega_{A}\otimes\Omega_{B}\subseteq\Omega_{A}\otimes_{max}\Omega_{B}$.
For the classical case ($A$ and $B$ classical) we will have
$\Omega_{A}\otimes_{min}\Omega_{B}=\Omega_{A}\otimes_{max}\Omega_{B}$.
For the quantum case we have the strict inclusions
$\Omega_{A}\otimes_{min}\Omega_{B}\subset\Omega_{A}\otimes\Omega_{B}\subset\Omega_{A}\otimes_{max}\Omega_{B}$.

\nd One can reasonably conceive of  a separable state in an
arbitrary COM as  one which may be written as a convex combination
of product states \cite{Barnum-Toner,Perinotti-2011}, i.e.

\begin{definition}\label{d:generalseparable}
A state $\omega\in\Omega_{A}\otimes\Omega_{B}$ will be called
\emph{separable} if there exist $p_{i}$,
$\omega^{i}_{A}\in\Omega_{A}$ and $\omega^{i}_{B}\in\Omega_{B}$ such
that

\begin{equation}
\omega=\sum_{i}p_{i}\omega^{i}_{A}\otimes\omega^{i}_{B}
\end{equation}

\end{definition}

\nd If $\omega\in\Omega_{A}\otimes\Omega_{B}$ but it is not
separable, we will  call it \emph{entangled}. {\sf Entangled
states exist only if $\Omega_{A}\otimes\Omega_{B}$ is strictly
greater than $\Omega_{A}\otimes_{min}\Omega_{B}$}.

\nd Using these constructions, marginal states can be defined as
follows \cite{Barnum-PRL}. Given a state
$\omega\in\Omega_{A}\otimes\Omega_{B}$, define

\begin{subequations}\label{e:marginalstates}

\begin{equation}
\omega_{A}(a) := \omega(a\otimes u_{B})
\end{equation}

\begin{equation}
\omega_{B}(b) := \omega(u_{A}\otimes b)
\end{equation}

\end{subequations}

\nd It is possible to show that the marginals of an entangled state
are necessarily mixed, while those of an unentangled pure state are
necessarily pure.

\vskip 2mm \nd These definitions are sufficient for a
generalization of entanglement to arbitrary COM's. In the
following section we review a geometrical construction whose
generalization yields an alternative conceptualization of the
entanglement-notion. The new view  turns out to be more general
than the one summarized above.

\section{Geometrical Characterization of
Entanglement Using Maps}\label{s:paperPRA}

\nd Let us now focus attention on quantum mechanics for the time
being. For a compound system represented by a Hilbert space
$\mathcal{H}$ (we restrict ourselves in what follows to a finite
dimension), $\mathcal{S}(\mathcal{H})$ is the convex hull of the set
of all product states. Let $\mathcal{C}$ be the convex set of
quantum states and $\mathcal{L}_{\mathcal{C}}$ the set of all convex
subsets of $\mathcal{C}$ (with analogous definitions of
$\mathcal{C}_i$ and $\mathcal{L}_{\mathcal{C}_i}$ for its
subsystems, $i=1,2$).

\subsection{Canonical Maps}\label{s:CanonicalMaps}

\nd We focus attention now in the specially important map $\Pi$

\begin{definition}\label{e:assignment}
$$\Pi:\mathcal{C}\longrightarrow\mathcal{C}$$
$$\rho\mapsto \rho^{A}\otimes\rho^{B}.$$
\end{definition}

\nd It is of the essence that product states
$\rho=\rho^{A}\otimes\rho^{B}$ not only satisfy

\begin{equation}\label{e:productproperty}
\Pi(\rho^{A}\otimes\rho^{B})=\rho^{A}\otimes\rho^{B},
\end{equation}
\noindent but  {\it are the only states which do satisfy
(\ref{e:productproperty})}. Partial traces are particular maps
defined between $\mathcal{C}$, $\mathcal{C}_{1}$, and
$\mathcal{C}_{2}$:

\begin{eqnarray}
&\mbox{tr}_{i}:\mathcal{C}\longrightarrow \mathcal{C}_{j}&\nonumber\\
&\rho\mapsto \mbox{tr}_{i}(\rho)&,
\end{eqnarray}

\noindent from which we can construct the induced maps $\tau_i$,
also very important for our present purposes, on
$\mathcal{L}_{\mathcal{C}}$, via the image of any subset
$C\subseteq\mathcal{C}$ under $\mbox{tr}_{i}$

\begin{eqnarray}\label{e:Taui}
&\tau_{i}:\mathcal{L}_{\mathcal{C}}\longrightarrow
\mathcal{L}_{\mathcal{C}_{i}}&\nonumber\\
&C\mapsto \mbox{tr}_{j}(C)&,
\end{eqnarray}

\noindent where for $i=1$ we take the partial trace with $j=2$ and
vice versa. In turn, we can define the product map

\begin{eqnarray} \label{e:Tau}
&\tau:\mathcal{L}_{\mathcal{C}}\longrightarrow\mathcal{L}_{\mathcal{C}_{1}}\times\mathcal{L}_{\mathcal{C}_{2}}&\nonumber\\
&C\mapsto(\tau_{1}(C),\tau_{2}(C))&.
\end{eqnarray}
 \noindent  \fbox{\parbox{2.9in}{This map  generalizes partial traces to convex subsets of
$\mathcal{C}$.}}

\nd  Given the convex subsets $C_1\subseteq\mathcal{C}_1$ and
$C_2\subseteq\mathcal{C}_2$ it is possible to define a product

\begin{definition}\label{d:tensorconvex}
Given the convex subsets $C_{1}\subseteq\mathcal{C}_{1}$ and
$C_{2}\subseteq\mathcal{C}_{2}$ we introduce now

\begin{equation}
C_1\otimes C_2:=\{\rho_{1}\otimes\rho_{2}\,|\,\rho_{1}\in
C_1,\rho_{2}\in C_2\}
\end{equation}
\end{definition}

\noindent Using this, we define the (for us all-important) map
$\Lambda$:

\begin{definition}\label{d:lambda}
$$\Lambda:\mathcal{L}_{\mathcal{C}_{1}}\times\mathcal{L}_{\mathcal{C}_{2}}\longrightarrow\mathcal{L}_{\mathcal{C}}$$
$$(C_{1},C_{2})\mapsto Conv(C_1\otimes C_2)$$
\end{definition}

\noindent where $Conv(\cdots)$ stands for \emph{convex hull} of a
given set. Applying $\Lambda$ to the particular case of the quantum
sets of states of the subsystems ($\mathcal{C}_{1}$ and
$\mathcal{C}_{2}$), one sees that Definitions \ref{d:tensorconvex}
and \ref{d:lambda} entail

\begin{equation}\label{e:LambdaC1C2}
\Lambda(\mathcal{C}_{1},\mathcal{C}_{2})=Conv(\mathcal{C}_{1}\otimes\mathcal{C}_{2})
\end{equation}

\noindent and this is nothing but

\begin{equation}\label{e:separablemixing}
\Lambda(\mathcal{C}_{1},\mathcal{C}_{2})=\mathcal{S}(\mathcal{H}),
\end{equation}
\noindent because $\mathcal{S}(\mathcal{H})$ is by definition (for
finite dimension) the convex hull of the set of all product states.

\subsection{Informational Invariance}

\noindent \fbox{\parbox{2.9in}{The map $\Lambda$ gives a precise
mathematical expression to the operation of making tensor products
and mixing, which has a clear physical meaning.}}

\nd  Let us elaborate: if it is possible to prepare in the
laboratory $A$ a given set of states $C_1$, it is reasonable to
assume that $C_1$ is convex, because if it is not, it is possible
to make it convex by recourse to classical algorithms (for
example, by tossing a biased coin, preparing one state or the
other according to the outcome, and then forgetting the outcome).
Same for the set $C_2$ in laboratory $B$. Then, it is possible
(without any recourse to non-classical interactions) to prepare
all product states of the form $\rho_{1}\otimes\rho_{2}$ with
$\rho_{1}\in C_1$ $\rho_{2}\in C_2$. Also, it is possible to
prepare all possible mixtures of such product states using a
classical algorithm of the type mentioned above. Now, this new set
of states is nothing but $\Lambda(C_1,C_2)$. Thus,
$\Lambda(C_1,C_2)$ is the maximal set of states which can be
generated without using non-classical correlations, given that the
set of states $C_1$ is available at laboratory $A$, and $C_2$ is
available in $B$.

\vskip 2mm \nd In particular, equation (\ref{e:separablemixing})
entails that {\it the set of all separable states of $\mathcal{C}$
is the image of the pair $(\mathcal{C}_{1},\mathcal{C}_{2})$ under
the map $\Lambda$}, i.e., all possible products and their mixtures
for the whole sets of states $\mathcal{C}_{1}$ and
$\mathcal{C}_{2}$.

\vskip 2mm  \nd Let us now turn to the function $\Lambda\circ\tau$
(the composition of $\tau$ with $\Lambda$)
\cite{Holik-Plastino-2011a}. For the special case of a convex set
formed by only one ``matrix'' (point) $\{\rho\}$ we have

\begin{equation}\label{e:lambdaenunrho}
\Lambda\circ\tau(\{\rho\})=\{\rho^{A}\otimes\rho^{B}\}
\end{equation}

\noindent which is completely equivalent to $\Pi$ (see Definition
\ref{e:assignment}), and thus satisfies an analogue of Equation
\eqref{e:productproperty}. Using this function it is possible to
derive a separability criterium in terms of properties of convex
sets that are polytopes \cite{Holik-Plastino-2011a}:

\nd \fbox{\parbox{2.9in}{
\begin{prop}\label{p:our criteria}
$\rho\in\mathcal{S}(\mathcal{H})$ if and only if there exists a
polytope $S_{\rho}$ such that $\rho\in S_{\rho}$ and
$\Lambda\circ\tau(S_{\rho})=S_{\rho}$.
\end{prop}}}

\nd Let us consider now the separability of pure states. Its
characterization in the bipartite instance is quite simple. We
assert that $\rho=|\psi\rangle\langle\psi|$ will be separable if and
only if it is a product of pure reduced states, i.e., if and only if
there exist $|\phi_{2}\rangle\in\mathcal{H}_{1}$ and
$|\phi_{2}\rangle\in\mathcal{H}_{2}$ such that
$|\psi\rangle=|\phi_{1}\rangle\otimes|\phi_{2}\rangle$. In
mathematical terms, this can be written as

\begin{eqnarray}\label{e:pureseparable}
& |\psi\rangle\langle\psi|\in\mathcal{S}(\mathcal{H})
\Leftrightarrow\Lambda\circ\tau(\{|\psi\rangle\langle\psi|\})=\{|\psi\rangle\langle\psi|\}\nonumber\\
&(\Leftrightarrow\Pi(|\psi\rangle\langle\psi|)=|\psi\rangle\langle\psi|).
\end{eqnarray}

\nd \fbox{\parbox{2.9in}{Equation \eqref{e:pureseparable} tells us
that a \emph{pure} state is separable, if and only if it remains
invariant under the function $\Lambda\circ\tau$ (or equivalently,
invariant under $\Pi$).}}

\nd  While this criterium is no longer valid for general mixed
states, the more general criterium \ref{p:our criteria} is
available for this case: a general \emph{mixed} state $\rho$ is
separable if and only if there exists a convex subset $S_{\rho}$
invariant under $\Lambda\circ\tau$. It is clear that the criterium
\ref{p:our criteria} is analogous to (\ref{e:pureseparable}),
being a generalization of it to convex subsets of $\mathcal{C}$,
with $\Lambda\circ\tau$ playing the role of the generalization of
$\Pi$. Thus, a generalization of the notion of product state for
convex sets can now be defined \cite{Holik-Plastino-2011a}

\begin{definition}
A convex subset $C\subseteq\mathcal{C}$ such that
$\Lambda\circ\tau(C)=C$ is called a \emph{convex separable subset}
(CSS) of $\mathcal{C}$.
\end{definition}

\nd Product states are limit cases of convex separable subsets (they
constitute the special case when the CSS has only one point)
\cite{Holik-Plastino-2011a}. CSS have the property of being
\emph{informational invariants} in the sense that the information
that they contain as probability spaces
\cite{Holik-Massri-Plastino-Zuberman} may be recovered via tensor
products and mixing of their (induced) reduced sub-states.

\vskip 2mm \noindent Let us turn now to a distinctive property of
$\Pi$. It is possible to prove that if $\Pi$ is applied twice  is
seen to be idempotent, i.e.,

\begin{equation}\label{e:omegacuadrado}
\Pi^{2}=\Pi.
\end{equation}

\nd and the same holds for $\Lambda\circ\tau$

\begin{equation}\label{e:lambdataucuadrado}
(\Lambda\circ\tau)^{2}=\Lambda\circ\tau.
\end{equation}

\noindent Consequently, the generalization of $\Pi$ satisfies an
equality equivalent to (\ref{e:omegacuadrado}).

\vskip 2mm \noindent An important remark is to be made. It is easy
to show that if we apply $\tau_i$ to $\mathcal{C}$, we obtain
$\mathcal{C}_i$. Thus, using Equation \eqref{e:separablemixing}, we
obtain

\begin{equation}
\Lambda\circ\tau(\mathcal{S}(\mathcal{H}))=\mathcal{S}(\mathcal{H})
\end{equation}

\noindent and thus, \emph{$\mathcal{S}(\mathcal{H})$ is itself an
informational invariant (a CSS), and in fact, the largest one}. As
we shall see in the following Sections, this fact can be gainfully
used to define separability and generalize the geometrical structure
of entanglement to arbitrary statistical theories. We shall also see
that the generalization of the properties of the functions
$\Lambda$, $\tau$ and $\Lambda\circ\tau$, allow us to see how to
define a huge family of entanglement measures in arbitrary COM's.

%%%%%%%%%%%%%%%%%%%%%%%%%%%%%%%%%%%%%%%%%%%%%%%% Bist hier

\section{Entanglement and separability in arbitrary convexity
models}\label{s:EntanglementGeneralization}

\nd In \cite{Beltrametti.Varadarajan-2000}, a general study of
extensions of convex operational models is presented. This general
framework includes compound systems. We will follow that paper's
approach to advance our entanglement-generalization, applicable to
arbitrary extensions of convexity models.

\subsection{Extensions of Convexity Models}\label{s:Varadarajan}

\nd Given two arbitrary convex operational models $\mathbf{A}$ and
$\mathbf{B}$ (see Section \ref{s:COMpreliminaries}) representing two
systems (they not necessarily possess the same underlying theory), a
morphism between them will be given by an affine map
$\phi:\Omega_{A}\rightarrow \Omega_{B}$ such that the affine dual
map $\varphi^{\ast}$ -defined by the functional
$\varphi^{\ast}(b):=b\circ\varphi$ (where ``$\circ$" denotes
composition)- maps the effects of $\mathbf{B}$ into effects of
$\mathbf{A}$ \cite{Beltrametti.Varadarajan-2000}.

\nd \fbox{\parbox{2.9in}{An affine map, is intuitively understood as
the canonical mathematical expression of a map preserving the convex
structure, which is the structure underlying all statistical
theories.}}

\nd A link between (or process from)) $\mathbf{A}$ - $\mathbf{B}$
will be represented by a morphism $\phi:\mathbf{A}\rightarrow
\mathbf{B}$ such that, for every state
$\alpha\in\Omega_{\mathbf{A}}$, $u_{\mathbf{B}}(\phi(\alpha))\leq 1$
(this is a normalization condition). If we want to study processes,
$u_{\mathbf{B}}(\phi(\alpha))$ will represent the probability that
the process represented by $\phi$ takes place. In this way,
morphisms can be used to represent links between systems (see next
paragraph), as for example, ``being a subsystem of", as well as
processes understood as general evolutions in time, continuous or
not.

\nd COM-extensions are studied in
\cite{Beltrametti.Varadarajan-2000}. Let us remember therefrom the
definition of the ``extension"-notion.

\nd \fbox{\parbox{2.9in}{A COM $\mathbf{\mathbf{C}}$ will be said
to be an extension of $\mathbf{A}$ if there exists a morphism
$\phi:\Omega_{\mathbf{C}}\rightarrow\Omega_{\mathbf{A}}$ which is
surjective.}}

\vskip 2mm \nd We emphasize the great generality of this
formulation: in the above definition of ``extension", almost all
possible conceivable cases are contained. A subsystem of a
classical or (quantal) system constitutes an example of an
extension in the above sense (it is the  morphism of the canonical
set-theoretical projection in the classical case, and of partial
trace in the quantum instance). Not only subsystems of a compound
system are captured by this notion of extension. Also limits
between theories, or coarse grained versions of a given theory,
may be considered --under this characterization-- as extensions.

\subsection{General Formal Setting}
\nd In order to look for a generalization of entanglement which
captures the results of previous Sections, we must look at triads of
COM's $\mathbf{C}$, $\mathbf{C_{1}}$ and $\mathbf{C_{2}}$, with
states spaces $\Omega_{\mathbf{C}}$, $\Omega_{\mathbf{C}_{1}}$, and
$\Omega_{\mathbf{C}_{2}}$, such that there exist two morphisms
(extension maps) $\phi_{1}$ and $\phi_{2}$ in such a way that
$\mathbf{C}$ be an extension of both $\mathbf{C_{1}}$ and
$\mathbf{C_{2}}$.

\nd It is clear that the product map $\phi=(\phi_{1},\phi_{2})$ may
be considered as the best candidate for a generalization of the map
$\tau$ (see Equation \eqref{e:Tau}). But in order to have adequate
generalizations of partial traces, i.e., in order to obtain
equivalence with the marginal states defined in
\ref{e:marginalstates}), we need an additional condition: for any
product state $a=a_1\otimes a_2$, we should have
$\phi(a)=(\phi_{1}(a_1),\phi_{2}(a_2))=(a_1,a_2)$, i.e., the
extension maps, when applied to a product state, must yield the
corresponding factors of the product, as partial traces do. Thus, we
give the following definition:

\begin{definition}
An  extension map $\phi=(\phi_{1},\phi_{2})$ will be called
\emph{a generalized partial trace} between COM's $\mathbf{C}$,
$\mathbf{C}_1$ and $\mathbf{C}_2$ if it satisfies

\begin{itemize}

\item $\phi_1$ and $\phi_2$ are  surjective morphisms between $\Omega_{\mathbf{C}}$,
$\Omega_{\mathbf{C}_1}$, and $\Omega_{\mathbf{C}_2}$.

\item For any product state $a=a_1\otimes a_2$,
$\phi(a)=(a_1,a_2)$.

\end{itemize}

\end{definition}

\nd and this is how the notion of marginal state defined in
\ref{e:marginalstates} can be recovered using extensions maps, a
much more general notion, in the sense that a particular extension
$\phi$ needs not to be a generalized partial trace as defined above.

\noindent If we want an analogue of $\Lambda$ (definition IV.3),
we must demand additional requirements as well. We will denote the
sets of convex subsets of $\Omega_{\mathbf{C}}$ and
$\Omega_{\mathbf{C}_{i}}$ ($i=1,2$) by $\mathcal{L}_{\mathbf{C}}$
and $\mathcal{L}_{\mathbf{C}_{i}}$, respectively. We are looking
for a map $\Psi$ with the following property. Once the extension
maps $\phi_{i}$ are fixed, $\Psi$ should map any pair of non-empty
convex subsets $(C_{1},C_{2})$ of
$\mathcal{L}_{\mathbf{C}_{1}}\times \mathcal{L}_{\mathbf{C}_{2}}$
into a non-empty convex subset $C$ of $\mathbf{C}$ with the
following {\it compatibility property}: for any $c\in
C:=\Psi(C_{1},C_{2})$, the extension maps must satisfy
$\phi_{1}(c)\in C_{1}$ and $\phi_{2}(c)\in C_{2}$. This condition
means that the image of $(C_1,C_2)$ under the map $\Psi$ is
compatible with the sub-states assigned by the extension maps
$\phi_1$ and $\phi_2$.

\nd As the maps $\phi_{i}$ are morphisms, it is possible to use them
to define canonically induced functions on convex subsets, and then
to map convex subsets of $\Omega_{\mathbf{C}}$ into convex subsets
of $\Omega_{\mathbf{C}_{i}}$, i.e., between
$\mathcal{L}_{\mathbf{C}}$ and $\mathcal{L}_{\mathbf{C}_{i}}$ (there
is an analogy with the earlier language involving $\tau_{i}$'s and
partial traces: we can make similar definitions as those of
Equations \eqref{e:Taui} and \eqref{e:Tau}). With some abuse of
notation we will keep calling these maps $\phi_{i}'s$, without undue
harm. Summing up, we will use the following definition:

\begin{definition}\label{d:CompoundPreInvariant}

\nd A triad $\mathbf{C}$, $\mathbf{C}_{1}$, and $\mathbf{C}_{2}$
will be called a \emph{compound system endowed with a
pre-informational invariance-structure} if

\begin{enumerate}

\item There exist morphisms $\phi_{1}$ and $\phi_{2}$ such that
$\mathbf{C}$ is an extension of $\mathbf{C}_{1}$ and
$\mathbf{C}_{2}$.

\item There exists also  a map $\Psi:\mathcal{L}_{\mathbf{C}_{1}}\times\mathcal{L}_{\mathbf{C}_{2}}\rightarrow\mathcal{L}_{\mathbf{C}}$ which
maps a pair of non-empty convex subsets
$(C_{1},C_{2})\in\mathcal{L}_{\mathbf{C}_1}\times\mathcal{L}_{\mathbf{C}_2}$
into a nonempty convex subset $C\in\mathcal{L}_{\mathbf{C}}$, such
that for every $c\in C$, $\phi(c)=(\phi_{1}(c),\phi_{2}(c))\in
C_{1}\times C_{2}$.

\end{enumerate}

\end{definition}

\nd Notice (again) that the morphisms $\phi_{i}$ may  not be,
necessarily, generalized partial traces. Most  physical systems of
interest satisfy these requirements. As we shall see below, all
essential features of entanglement can be recovered using these
canonical maps between state spaces.

\nd The function $\Lambda$ (defined in \ref{d:lambda}) can be
naturally generalized to an arbitrary compound system as follows.
Given operational models $\mathbf{A}$, $\mathbf{B},$ and
$\mathbf{C}$, let $\mathcal{L}_{\mathbf{A}}$,
$\mathcal{L}_{\mathbf{B}},$ and $\mathcal{L}_{\mathbf{C}}$ be the
sets of convex subsets of $\Omega_{\mathbf{A}}$
$\Omega_{\mathbf{B}}$, and $\Omega_{\mathbf{C}}$, respectively, one
defines

\begin{definition}\label{d:widetilde}
$$\widetilde{\Lambda}: \mathcal{L}_{\Omega_{A}}\times
\mathcal{L}_{\Omega_{B}}\longrightarrow \mathcal{L}_{\Omega_{C}}$$
$$\widetilde{\Lambda}(C_{1},C_{2})\mapsto Conv(C_{1}\otimes C_{2}).$$
\end{definition}

\nd where $C_{1}\otimes C_{2}$ is defined as in \ref{d:tensorconvex}
and $Conv(\ldots)$ stands again for convex closure. It is easy to
check that the function defined by \ref{d:widetilde} represents a
particular case of a function of the type $\Psi$ (Definition
\ref{d:CompoundPreInvariant}). Notice that the functions $\Psi$ may
include more general examples, i.e, there are several forms of going
up from the subsystems to the system. For example, we may take

\begin{equation}
\Psi(C_{1},C_{2})=\phi_{1}^{-1}(C_{1})\cap\phi_{2}^{-1}(C_{2}),
\end{equation}

\nd (which in the quantum realm would correspond to
$\Psi(C_{1},C_{2})=\mbox{tr}_{1}^{-1}(C_{1})\cap\mbox{tr}_{2}^{-1}(C_{2})$).
If $C_{1}=\{\rho_{1}\}$ $C_{2}=\{\rho_{2}\}$, the function $\Psi$
thus defined yields a convex set of states which may be global ones,
compatible with given reduced states $\rho_{1}$ and $\rho_{2}$. It
should also be clear that a function $\Psi$ different from
$\widetilde{\Lambda}$ will arise in a model in which the extension
contains a third system (apart from $\mathbf{C}_{1}$ and
$\mathbf{C}_{2}$). \vskip 2mm

\nd Thus, we see that the definitions involved in
\ref{d:CompoundPreInvariant} are much more general than partial
traces and the $\Lambda$-map. In this sense, any new construction
that we define below which uses such functions, contains the usual
examples as particular cases.

\vskip 2mm \nd Before going on,  remark that the constructions
presented here represent a general setting for COM's. In this
setting, systems are represented as COM's with a given geometry and
the theory may depend critically on the specific choice of the maps
$\phi$ and $\Psi$. This choice may represent i) a structural feature
of the theory, as is, for example, the case of partial traces in
$QM$ (which link states of the system with states of the
subsystems), or ii) a theoretical aspect that we want investigate in
some detail (as for example, the problem of which global states are
compatible with two given reduced states of the subsystems mentioned
above). Once these maps and the geometry of the convex sets of
states (and observables) are specified, the formal setting is ready
for defining ``entanglement", informational invariance, and
entanglement measures.

\subsection{Generalized Entanglement}

\nd The extension $\widetilde{\Lambda}$ of the function $\Lambda$
to arbitrary statistical models, together with the notion of
generalized partial traces, allow for the extension of the notions
of informational invariance and CSS to any COM

\begin{definition}
A convex subset $C$ of the set of states $\Omega$ of a compound
statistical system $\mathbf{C}$ consisting of
$\mathbf{C}_{1}-\mathbf{C}_{2}$ and endowed with i) a generalized
partial trace $\phi$ and ii) the up-function $\widetilde{\Lambda}$
will be called a CSS if it satisfies

\begin{equation}
\widetilde{\Lambda}\circ\phi(C)=C.
\end{equation}

\end{definition}

\nd For finite dimension, using Carath\'{e}odory's theorem it is
also possible to show that if $\phi$ is a generalized partial
trace, a state $\rho$ of an arbitrary physical system may  be
appropriately  called separable, in the sense of definition
\ref{d:generalseparable}, if and only if there exists a CSS $C$
(e.g., such that $\widetilde{\Lambda}\circ\phi(C)=C$) such that
$\rho\in C$. The demonstration of this fact is analogous to that
of \ref{p:our criteria} \cite{Holik-Plastino-2011a}. Note that in
order that an equivalence with definition \ref{d:generalseparable}
may hold we must use $\widetilde{\Lambda}\circ\phi$ in the
definition of informational invariance (and not the more general
$\Psi\circ\phi$) with $\phi$ a generalized partial trace.  With
these constructions at hand, let us restrict ourselves, for the
sake of simplicity, to compound systems with only two subsystems
and {\it look for a generalization of the entanglement and
separability notions}. \vskip 2mm

\nd It should now be  clear that the analogues of the maps
$\Lambda$ and $\tau$ are $\widetilde{\Lambda}$ and $\phi$,
respectively. An important remark needs to be stated at this
point. If we have a classical compound $\mathbf{C}$ system, with
subsystems $\mathbf{C}_1$ and $\mathbf{C}_2$, then it is easy to
show that \emph{the whole set of states $\Omega$ is an
informational invariant}. This means that we have the following
proposition

\nd \fbox{\parbox{2.9in}{
\begin{prop}
If a system $\mathbf{C}$ with state space $\Omega$, formed by
subsystems $\mathbf{C}_1$ and $\mathbf{C}_2$ is classical, then
$\widetilde{\Lambda}\circ\phi(\Omega)=\Omega$.
\end{prop} This  proposition allows us to characterize
classicality as a special case of informational invariance.}}

\nd Note that informational invariance does not imply classicality:
the state space could be a CSS but not a simplex.

\nd \fbox{\parbox{2.9in}{ Any system for which its state space is
not information-invariant will exhibit entanglement.}}

\nd  \emph{In general, it will be reasonable to define the set of
separable states as the largest informational invariant subset}. In
particular, if separability is defined as in
\ref{d:generalseparable}, any state $\omega$ which does not belongs
to this maximally invariant subset (which is the set of separable
states as defined in \ref{d:generalseparable}), will satisfy
$\widetilde{\Lambda}\circ\tau(\omega)\neq\omega$. But it is
important to remark that a more general notion of non separability
will be given by the condition $\Psi\circ\tau(\omega)\neq\omega$.

\nd Thus, given a system which is an extension of two other
systems, an alternative definition/axiomatization of an
entanglement structure can be given by imposing conditions on the
maps $\phi$ and $\Psi$ as follows:

\begin{definition}\label{d:entanglementGeneralized}
Given a \emph{two component compound system endowed with a
pre-informational invariance structure} $\mathbf{C}$, formed by
$\mathbf{C}_{1}$, and $\mathbf{C}_{2}$, with up-map $\Psi$ and a
down-map $\phi$, then

\begin{enumerate}

\item A state $c\in\mathbf{C}$ will be called a \emph{non-product state} if
$\Psi\circ\phi(\{c\})\neq\{c\}$. Otherwise, it will be called a
\emph{product state}.

\item For an \emph{invariant convex subset} $C$ one has
$C\in\mathcal{L}_{\mathbf{C}}$, such that $\Psi\circ\phi(C)=C$.

\item If there exist a largest (in the sense of inclusion) invariant subset, we will denote it by
$\mathcal{S}(\mathbf{C})$.

\item A \emph{two-components compound system} for
which

\begin{itemize}

\item  there exists
$\mathcal{S}(\mathbf{C})$ and

\item strict inclusion in
$\mathbf{C}$ is guaranteed,

\end{itemize} will be said to be an \emph{entanglement operational
model}.

\item In an \emph{entanglement operational model} a state $c$ which satisfies
$c\notin\mathcal{S}(\mathbf{C})$ will be said to be entangled.

\end{enumerate}

\end{definition}

\nd \emph{It is clear that using these constructions we can export
the quantum entanglement structure to a wide class of COM's, and for
that reason, to many new statistical physics' systems. And this is
done by imposing conditions on very general notions, such as maps
between operational models.}

\nd If in the above definition we take $\Psi$ to be
$\widetilde{\Lambda}$ and $\phi$ a generalized partial trace,
entanglement is thus defined in terms of informational invariance.
It should be clear also that quantum mechanics is the best example
for entanglement, and that all states in classical mechanics are
separable. Remark that the properties of a two-components system
will depend, in a strong sense, on the choice of the functions
$\Psi$ and $\phi$. These should be selected as the canonical ones,
i.e., the ones which are somehow natural for the physics of the
problem under study.

\nd Nevertheless, we remark that nothing prevents us from making
more general choices for practical purposes. Then, we can also
``postulate" a generalized separability criterium (having
 a different ``content" than the one which uses
$\widetilde{\Lambda}$) that is not necessarily equivalent to the one
of  definition \ref{d:generalseparable}) and contains it as a
special case:

\begin{definition}
\item A state $c\in\mathbf{C}$ in an \emph{entanglement operational model}
is said to be separable iff there exists
$C\subseteq\mathcal{S}(\mathbf{C})$ containing $c$ such that
$\Psi\circ\phi(C)=C$.
\end{definition}

\nd Note that any general definition of the convex invariant subsets
can be formulated via the particular choice of the all-important
functions $\phi$ and $\Psi$. These constructions may be useful to
develop and search for generalizations/corrections of/to quantum
mechanics and for the study of quantum entanglement in theories of a
more general character than quantum mechanics. Our constructions
constitute a valid alternative to others that one can find in the
literature. An interesting open problem would be that of finding the
way in which we can express the violation of Bell's inequalities
using our present approach.

\subsection{Generalized Entanglement Measures}

\nd The constructions erected in previous sections give us a point
of view that suggests in clear fashion just how to generalize a
certain family of entanglement measures analogous to the
Schlienz-Mahler ones \cite{Ent95,Holik-Plastino-2011a}. Given that a
state $c$ will be entangled iff $\Psi\circ\phi(\{c\})\neq \{c\}$, it
is tempting to regard the difference between $\Psi\circ\phi(\{c\})$
and $\{c\}$ as a measure of entanglement. For the simple case in
which $\Psi\circ\phi(\rho)$ has only one element (as is the case if
$\Psi=\widetilde{\Lambda}$), we define (with some abuse of notation
in avoiding the set theoretical ``$\{\ldots\}$" symbols):

\begin{equation}\label{e:ExtensionMeasurement}
G(\rho):=\|H(\Psi\circ\phi(\rho)-\rho)\|,
\end{equation}

\nd with $H$ and $\|\ldots\|$ a convenient function and norm,
respectively. Thus, our construction includes a generalization of
a family of quantitative measures of entanglement for arbitrary
statistical models. One of the main advantages of this approach is
that it provides a completely geometrical formulation of
entanglement measures. For the quantum case, and taking
$\Psi=\Lambda$ and
$\phi=(\mbox{tr}_1(\ldots),\mbox{tr}_2(\ldots))$ the family
(\ref{e:ExtensionMeasurement}) adopts the form

\begin{equation}\label{e:SM}
SM(\rho)=\|F(\rho^{A}\otimes\rho^{B}-\rho)\|
\end{equation}

\noindent with $F$ and $\|\ldots\|$ a convenient function and norm,
respectively. It can be shown that they are computable and if $F$
and $\|\ldots\|$ are suitably chosen, they provide entanglement
criteria as strong as the celebrated \emph{Partial Transpose} one
(one of the strongest computable ones)
\cite{Abascal2007,Bjork-2007a,Bjork-2007b,Zhangh-2008}.

\nd Equation \ref{e:SM} may be reexpressed  as follows:

\nd \fbox{\parbox{2.9in}{ \emph{Given a state $\rho$, make the
tensor product of its partial traces (i.e., apply the map $\Pi$
defined in \ref{e:assignment}), compute an specified function of
their difference, and take the norm.}}}

\section{Conclusions}\label{s:conclusions}

%&/
\nd We have worked out our generalizations of some important
quantum mechanics' features  via the ``reciprocal" maps

\begin{itemize}
\item
\begin{eqnarray} \label{tautau}
&\tau:\mathcal{L}_{\mathcal{C}}\longrightarrow\mathcal{L}_{\mathcal{C}_{1}}\times\mathcal{L}_{\mathcal{C}_{2}}&\nonumber\\
&C\mapsto(\tau_{1}(C),\tau_{2}(C))&
\end{eqnarray}
\noindent which generalizes partial traces to convex subsets of
$\mathcal{C}$.

\item $$\Lambda:\mathcal{L}_{\mathcal{C}_{1}}\times\mathcal{L}_{\mathcal{C}_{2}}\longrightarrow\mathcal{L}_{\mathcal{C}}$$
$$(C_{1},C_{2})\mapsto Conv(C_1\otimes C_2),$$
\noindent where $Conv(\cdots)$ stands for \emph{convex hull}.
Applying $\Lambda$ to the particular case of ordinary quantum sets
of states of two subsystems ($\mathcal{C}_{1}$ and
$\mathcal{C}_{2}$), one sees that
\begin{equation}\label{e:separablemixingzzz}
\Lambda(\mathcal{C}_{1},\mathcal{C}_{2})=\mathcal{S}(\mathcal{H}),
\end{equation}
\noindent the set of all separable states, i.e., for finite
dimension, the convex hull of the set of all product states.

\end{itemize}

\nd We can summarize our results as follows:

\begin{itemize}

\item We provided a generalization of some geometrical properties
of entanglement to any statistical theory via the COM approach. This
is done by generalizing a previously discovered geometrical
structure (see \cite{Holik-Plastino-2011a}). The generalization is
achieved by imposing conditions between very general maps defined
between convex operational models, enriching the approach presented
in \cite{Beltrametti.Varadarajan-2000}. Although there is a standard
way in which entanglement may be generalized (provided by definition
\ref{d:generalseparable}), our approach is different and poses the
emphasis on the maps mentioned above. Our present framework possess
the advantage of being describable in purely geometrical terms.
Because of the great generality of the COM approach, these
constructions hold for all statistical theories.

\item In particular, we presented the extension of the maps $\Lambda$ and $\tau$
($\Psi$ and $\phi$, respectively) to {\it arbitrary} statistical
models. We showed that it is possible to generalize $\Lambda$ in any
COM with the map $\widetilde{\Lambda}$ [Cf.
Definition(\eqref{d:widetilde})].

\item The alternative perspective provided by these generalizations allows us to define

(1) new families of entanglement measures, valid for arbitrary
statistical models [Cf. Eq. \eqref{e:ExtensionMeasurement}] which
are based on the Schlienz-Mahler one \cite{Ent95}, and \newline (2)
also yields appropriate extensions of the notions of informational
invariance and convex separable subsets (CSS) to any arbitrary COM.

\end{itemize}

\appendix
\section{Quantal effects}\label{s:COMapproach}

\nd In modeling probabilistic operational theories one associates to
any probabilistic system a triplet $(X,\Sigma,p)$, where
\begin{enumerate}
\item $\Sigma$ represents the set of states of the system,
\item $X$ is the set of possible measurement outcomes, and
\item $p:X\times \Sigma\mapsto [0,1]$ assigns to each outcome $x\in
X$ and state $s\in\Sigma$ a probability $p(x,s)$ of $x$ to occur
if the system is in the state $s$.  \item   If we fix $s$ we
obtain the mapping $s\mapsto p(\cdot,s)$ from $\Sigma\rightarrow
[0,1]^{X}$.
\end{enumerate}

\nd Note that
\begin{itemize}
\item This  identifies all the states of $\Sigma$ with maps.
\item Considering their closed convex hull, we obtain the set
$\Omega$ of possible probabilistic mixtures (represented
mathematically by convex combinations) of states in $\Sigma$.
\item In this way one also obtains, for any outcome $x\in X$, an
affine evaluation-functional $f_{x}:\Omega\rightarrow [0,1]$,
given by $f_{x}(\alpha)=\alpha(x)$ for all $\alpha\in \Omega$.
\item  More generally, any affine functional $f:\Omega\rightarrow
[0,1]$ may be regarded as representing a measurement outcome and
thus use $f(\alpha)$ to represent the probability for that outcome
in state $\alpha$.
\end{itemize}

\nd  For the special case of quantum mechanics, the set of all
affine functionals so-defined are called effects. They form an
algebra (known as the \emph{effect algebra}) and represent
generalized measurements (unsharp, as opposed to sharp measures
defined by projection valued measures). The specifical form of an
effect in quantum mechanics is as follows. A generalized
observable or \emph{positive operator valued measure} (POVM) will
be represented by a mapping

\begin{subequations}

\begin{equation}
E:B(\mathcal{R})\rightarrow\mathcal{B}(\mathcal{H})
\end{equation}

\noindent such that

\begin{equation}
E(\mathcal{R})=\mathbf{1}
\end{equation}

\begin{equation}
E(B)\geq 0, \,\,\mbox{for any}\,\, B\in B(\mathcal{R})
\end{equation}

\noindent and for any disjoint family $\{B_{j}\}$

\begin{equation}
E(\cup_{j}(B_{j}))=\sum_{j}E(B_{j}).
\end{equation}

\end{subequations}

\noindent The first condition means that $E$ is normalized to
unity, the second one that $E$ maps any Borel set B to a positive
operator, and the third one that $E$ is $\sigma$-additive with
respect to the weak operator topology. In this way, a generalized
POVM can be used to define a family of affine functionals on the
state space $\mathcal{C}$ (which corresponds to $\Omega$ in the
general probabilistic setting) of quantum mechanics as follows

\begin{subequations}

\begin{equation}
E(B):\mathcal{C}\rightarrow [0,1]
\end{equation}

\begin{equation}
\rho\mapsto \mbox{tr}(E\rho)
\end{equation}

\end{subequations}

\noindent Positive operators $E(B)$ which satisfy $0\leq
E\leq\mathbf{1}$ are called effects (which form an \emph{effect
algebra}. Let us denote by $\mathrm{E}(\mathcal{H})$ the set of all
effects.\vskip 3mm

\nd {\sf Indeed, a POVM is a measure whose values are non-negative
self-adjoint operators on a Hilbert space. It is the most general
formulation of a measurement in the theory of quantum physics}.
\vskip 3mm \nd A rough analogy would consider that a POVM is to a
projective measurement what a density matrix is to a pure state.
Density matrices can describe part of a larger system that is in a
pure state (purification of quantum state); analogously, POVMs on a
physical system can describe the effect of a projective measurement
performed on a larger system. Another, slightly different way to
define them is as follows:

\nd Let $(X, M)$ be measurable space; i.e., $M$ is a
$\sigma-$algebra of subsets of $X$. A POVM is a function $F$ defined
on $M$ whose values are bounded non-negative self-adjoint operators
on a Hilbert space $\mathcal{H}$ such that $F(X) = I_H$ (identity)
and for every i) $\xi \in  \mathcal{H}$ and ii) projector $P=
|\psi\rangle\langle  \psi|;\,\, |\psi\rangle \in \mathcal{H}$,  $P
\rightarrow\, \langle F(P)\xi \vert  \xi \rangle$ is a  non-negative
countably additive measure on $M$. This definition should be
contrasted with that for the projection-valued measure, which is
very similar, except that, in the projection-valued measure, the
$F$s are required to be projection operators.

\vskip1truecm

\noindent {\bf Acknowledgements} \noindent This work was partially
supported by the following grants:  i)  project PIP1177 of CONICET
(Argentina) and ii)   project FIS2008-00781/FIS (MICINN) and FEDER
(EU) (Spain, EU).

\end{document}